\pdfoutput=1
\documentclass{article}

\PassOptionsToPackage{numbers, compress}{natbib}
\usepackage[preprint]{nips_2018}

\usepackage[utf8]{inputenc} % allow utf-8 input
\usepackage[T1]{fontenc}    % use 8-bit T1 fonts
\usepackage{hyperref}       % hyperlinks
\usepackage{url}            % simple URL typesetting
\usepackage{booktabs}       % professional-quality tables
\usepackage{amsfonts}       % blackboard math symbols
\usepackage{nicefrac}       % compact symbols for 1/2, etc.
\usepackage{microtype}      % microtypography

\usepackage{algorithm}
\usepackage[noend]{algorithmic}
\usepackage{mathtools}
\usepackage{amsmath}
\usepackage{amsthm}
\usepackage{amssymb}
\usepackage{enumitem}
\newtheorem{definition}{Definition}

\usepackage{mathtools}
\mathtoolsset{centercolon}
\newcommand{\defeq}{\mathrel{:\mkern-0.25mu=}}

\newcommand{\bbI}{\ensuremath{\mathbb{I}}}

\newcommand{\bbR}{\ensuremath{\mathbb{R}}}

\newcommand{\cD}{\ensuremath{\mathcal{D}}}

\newcommand{\cX}{\ensuremath{\mathcal{X}}}
\newcommand{\cY}{\ensuremath{\mathcal{Y}}}

\newcommand{\cfrp}{CFR$^+$}
\newcommand{\rmp}{RM$^+$}

\newcommand{\egt}{EGT}
\newcommand{\efg}{EFG}

\def\epsilonsad{{\epsilon_{\hbox{\scriptsize\rm sad}}}}

\def\argmin{\mathop{\hbox{\rm argmin}}}

\newcommand{\be}{\begin{eqnarray}}
\newcommand{\ee}[1]{\label{#1}\end{eqnarray}}

\newcommand{\ese}{\end{eqnarray*}}
\newcommand{\bse}{\begin{eqnarray*}}
\def\beq{\begin{equation}}
\def\eeq{\end{equation}}

\def\fnote#1{\footnote}

\def\*{{{\LARGE\bf $^*$}}}

\def\R{{\mathbb{R}}}

\def\cD{{\cal D}}

\def\cX{{\cal X}}
\def\cY{{\cal Y}}

\def\argmin{\mathop{\rm argmin}}

\def\log{\mathop{{\rm log}}}

\usepackage{caption}
\title{Solving Large Sequential Games with the Excessive Gap Technique}

\author{
  Christian Kroer, Gabriele Farina, and Tuomas Sandholm \\
  Department of Computer Science\\
  Carnegie Mellon University\\
  Pittsburgh, PA 15213 \\
  \texttt{\tt{\{ckroer,gfarina,sandholm\}}@cs.cmu.edu} \\
}

\begin{document}

\maketitle

\begin{abstract}
	There has been tremendous recent progress on equilibrium-finding algorithms for
zero-sum imperfect-information extensive-form games, but there has been a
puzzling gap between theory and practice. \emph{First-order methods} have
significantly better theoretical convergence rates than any
\emph{counterfactual-regret minimization (CFR)} variant. Despite this, CFR
variants have been favored in practice. Experiments with first-order methods
have only been conducted on small- and medium-sized games because those methods
are complicated to implement in this setting, and because CFR variants have been
enhanced extensively for over a decade they perform well in practice. In this
paper we show that a particular first-order method, a state-of-the-art variant
of the \emph{excessive gap technique}---instantiated with the \emph{dilated
  entropy distance function}---can efficiently solve large real-world problems
competitively with CFR and its variants. We show this on large endgames
encountered by the \emph{Libratus} poker AI, which recently beat top human poker
specialist professionals at no-limit Texas hold'em. We show experimental results
on our variant of the excessive gap technique as well as a prior version. We
introduce a numerically friendly implementation of the smoothed best response
computation associated with first-order methods for extensive-form game solving.
We present, to our knowledge, the first GPU implementation of a first-order
method for extensive-form games. We present comparisons of several excessive gap
technique and CFR variants.
%%% Local Variables:
%%% mode: latex
%%% TeX-master: "../nips18/gpu_egt_river_nips18"
%%% End:

\end{abstract}

\section{Introduction}
Two-player zero-sum extensive-form games (EFGs) are a general representation
that enables one to model a myriad of  settings ranging from security to
business to military to recreational. The \emph{Nash equilibrium }solution
concept~\citep{Nash50:Equilibrium} prescribes a sound notion of rational play
for this setting. It is also robust in this class of game: if the opponent plays
some other strategy than an equilibrium strategy, that can only help us.

There has been tremendous recent progress on equilibrium-finding algorithms for
extensive-form zero-sum games. However, there has been a vexing gap between the
theory and practice of equilibrium-finding algorithms. In this paper we will
help close that gap.

It is well-known that the strategy spaces of an extensive-form game can be
transformed into convex polytopes that allow a bilinear saddle-point formulation
(BSPP) of the Nash equilibrium problem as
follows~\citep{Romanovskii62:Reduction,Stengel96:Efficient,Koller96:Efficient}.
\begin{equation}
  \label{eq:sequence_form_objective} \min_{x \in \cX} \max_{y \in \cY} \langle
x, Ay \rangle = \max_{y \in \cY} \min_{x \in \cX} \langle x,Ay \rangle
\end{equation}

Problem~\eqref{eq:sequence_form_objective} can be solved in a number of ways.
Early on, \citet{Stengel96:Efficient} showed that it can be solved with a linear
program (LP)---by taking the dual of the optimization problem faced by one
player (say the $y$ player) when holding the strategy of the $x$ player fixed,
and injecting the primal $x$-player constraints into the dual LP\@. This approach
was used in early work on extensive-form game solving, up to games of size
$10^5$~\citep{Koller97:Representations}. \citet{Gilpin07:Lossless} coupled it
with lossless abstraction in order to solve Rhode Island hold'em which has
$10^9$ nodes in the game tree. Since then, LP approaches have fallen out of
favor.  The LP is often too large to fit in memory, and even when it does fit
the iterations of the simplex or interior-point methods used to solve the LP
take too long---even if only modest accuracy is required. 

Instead, modern work on solving this game class in the large focuses on
iterative methods that converge to a Nash equilibrium in the limit. Two types of
algorithms have been popular in particular: regret-minimization algorithms based
on \emph{counterfactual regret minimization
(CFR)}~\citep{Zinkevich07:Regret,Lanctot09:Monte,Bowling15:Heads,Brown15:Hierarchical,Moravvcik17:DeepStack,Brown17:Superhuman},
and \emph{first-order methods (FOMs)} based on combining a fast \emph{bilinear
saddle-point problem (BSPP)} solver such as the \emph{excessive gap technique
(EGT)}~\citep{Nesterov05:Excessive} with an appropriate
\emph{distance-generating function (DGF)} for EFG
strategies~\citep{Hoda10:Smoothing,Kroer15:Faster,Kroer17:Theoreticala,Kroer17:Smoothing}.

The CFR family has been most popular in practice so far. The {\cfrp}
variant~\citep{Tammelin15:Solving}  was used to near-optimally solve heads-up
limit Texas hold'em~\citep{Bowling15:Heads}, a game that has $10^{13}$ decision
points after lossless abstraction. {\cfrp} was also used for subgame solving in
two recent man-machine competitions where AIs beat human poker pros at no-limit
Texas hold'em~\citep{Moravvcik17:DeepStack,Brown17:Superhuman}---a game that has
$10^{161}$ decision points (before abstraction)~\citep{Johanson13:Measuring}. A
variant of CFR was also used to compute the whole-game strategy (aka.
``blueprint'' strategy) for \emph{Libratus}, an AI that beat top specialist pros
at that game~\citep{Brown17:Superhuman}.

CFR-based algorithms converge at a rate of $\frac{1}{\sqrt{T}}$, whereas some
algorithms based on FOMs converge at a rate of $\frac{1}{T}$. Despite this
theoretically superior convergence rate, FOMs have had relatively little
adoption in practice. Comparisons of CFR-based algorithms and FOMs were
conducted by \citet{Kroer15:Faster} and \citet{Kroer17:Theoreticala}, where they
found that a heuristic variant of EGT instantiated with an appropriate distance
measure is superior to CFR regret matching (RM) and CFR with regret-matching$^+$
({\rmp}) for small-to-medium-sized games. 

In this paper, we present the first experiments on a large game---a real game
played by humans---showing that an aggressive variant of EGT instantiated with
the DGF of \citet{Kroer17:Theoreticala} is competitive with the CFR family in
practice. It outperforms CFR with {\rmp}, although {\cfrp} is still slightly
faster. This is the first time that a FOM has been shown superior to any CFR
variant on a real-world problem. We show this on subgames encountered by
\emph{Libratus}. The \emph{Libratus} agent solved an abstraction of the full
game of no-limit Texas hold'em ahead of time in order to obtain a ``blueprint''
strategy. During play, \emph{Libratus} then refined this blueprint strategy by
solving subgames with significantly more detailed abstractions in real
time~\citep{Brown17:Superhuman,Brown17:Safe}.
% , in particular the
%\emph{turn} (the betting round before the last public card is dealt) and
%\emph{river} (the betting round after the last card is dealt). 
Our experiments are on solving endgames encountered by \emph{Libratus} in the
beginning of the fourth (``river'' in poker lingo) betting round, with the full
fine-grained abstraction actually used by \emph{Libratus}. This abstraction has
no abstraction of cards, that is, the model captures all aspects of the cards.
There is abstraction of bet sizes to keep the branching factor reasonable; in
our experiments we use the exact full fine-grained betting abstraction that was
used by \emph{Libratus}. Thus we show that it is possible to get the
theoretically superior guarantee of FOMs while also getting strong practical
performance.

In order to make our approach practical, we introduce a number of practical
techniques for running FOMs on EFGs. In particular, we derive efficient and
numerically friendly expressions for the \emph{smoothed-best response (SBR)} and
\emph{prox mapping}, two optimization subproblems that EGT solves at every
iteration. Furthermore, we introduce a GPU-based variant of these operations
which allows us to parallelize EGT iterations.

We show experiments for several variants of both EGT and CFR\@. For EGT, we
consider two practical variants, one that has the initial smoothing parameter
set optimistically, and one that additionally performs aggressive stepsizing.
For CFR, we show experimental results for CFR with RM, \rmp, and {\cfrp} (i.e.,
CFR with linear averaging and \rmp). We will describe these variants in detail
in the body of the paper. We conducted all the experiments on parallelized GPU
code.

%%% Local Variables:
%%% mode: latex
%%% TeX-master: "../nips18/gpu_egt_river_nips18"
%%% End:

% related work (maybe just in intro)
% EFGs
% EGT

\section{Bilinear Saddle-Point Problems}

The computation of a Nash equilibrium in a zero-sum imperfect-information EFG
can be formulated as the following bilinear saddle-point problem:
\begin{align}
  \label{eq:bspp}
  \min_{x\in\cX}\max_{y\in\cY} \langle x, Ay \rangle = \max_{y\in\cY}\min_{x\in\cX} \langle x, Ay \rangle,
\end{align}
% where $\phi(x,y) = v + \langle a_1, x \rangle + \langle a_2, y + \rangle\langle x,
% Ay \rangle$ and
where $\cX,\cY$ are convex, compact sets in Euclidean spaces
$E_x,E_y$.
% We let $\cZ=\cX\times \cY$ denote the Cartesian product of $\cX,\cY$
% so that $\phi(x,y):\cZ \rightarrow \bbR$. 
% For the case of EFGs (\ref{eq:bspp})
% takes the form of (\ref{eq:sequence_form_objective}) so that $\phi(x,y)=\langle
% x, Ay \rangle$ where
$A$ is the sequence-form payoff matrix and $\cX,\cY$ are
the sequence-form strategy spaces of Player 1 and 2, respectively.

Several FOMs with attractive convergence properties have been introduced for
BSPPs~\cite{Nesterov05:Smooth,Nesterov05:Excessive,Nemirovski04:Prox,Chambolle11:First}.
These methods rely on having some appropriate distance measure over $\cX$ and
$\cY$, called a \emph{distance-generating function} (DGF). Generally, FOMs use
the DGF to choose steps: given a gradient and a scalar stepsize, a FOM moves in
the negative gradient direction by finding the point that minimizes the sum of
the gradient and of the DGF evaluated at the new point. In other words, the next
step can be found by solving a regularized optimization problem, where long
gradient steps are discouraged by the DGF\@. For EGT on EFGs, the DGF can be
interpreted as a smoothing function applied to the best-response problems faced
by the players.

\begin{definition}
  A distance-generating function for $\cX$ is a function $d(x):\cX \rightarrow
  \bbR$ which is convex and continuous on $\cX$, admits continuous selection of
  subgradients on the set $\cX^\circ=\left\{ x\in\cX: \partial d(x) \ne \emptyset
  \right\}$, and has strong convexity modulus $\varphi$ w.r.t. $\|\cdot\|$.
  Distance-generating functions for $\cY$ are defined analogously.
\end{definition}

% Given a twice differentiable function $f$, we let $\nabla^2f(z)$ denote its Hessian at $z$. Our analysis is based on the following sufficient condition for strong convexity of a twice differentiable function:
% \begin{fact} \label{fac:strong_convexity_hessian}
%   A twice-differentiable function $f$ is strongly convex with modulus $\varphi$ with respect to a norm $\|\cdot\|$ on nonempty convex set $C\subset\bbR^n$ if
% $
%   h^\top \nabla^2f(z) h \geq \varphi\|h\|,\ \forall h\in\bbR^n, z\in C^\circ.
% $
% \end{fact}

% For a twice-differentiable function, a sufficient condition for strong convexity with respect to a norm $\|\cdot\|$ is
% $
%   h^\top \nabla^2d(x) h \geq \varphi\|h\|,
% $ for all $h\in\bbR^n$ and $x\in \cX^\circ$.

Given DGFs $d_{\cX},d_{\cY}$ for $\cX,\cY$ with strong convexity moduli
$\varphi_{\cX}$ and $\varphi_{\cY}$ respectively, we now describe
EGT~\cite{Nesterov05:Excessive} applied to \eqref{eq:sequence_form_objective}.
EGT forms two smoothed functions using the DGFs
\begin{align}
\vspace{-1mm}
  f_{\mu_y}(x) = \max_{y\in \cY} \langle x, Ay \rangle - \mu_y d_\cY,\qquad
  \phi_{\mu_x}(y) = \min_{x\in \cX} \langle x, Ay \rangle + \mu_x d_\cX .\label{eq:smoothed_functions}
\vspace{-1mm}
\end{align}
These functions are smoothed approximations to the optimization problem faced by
the $x$ and $y$ player, respectively. The scalars $\mu_x,\mu_y>0$ are smoothness
parameters denoting the amount of smoothing applied. Let $y_{\mu_y}(x)$ and
$x_{\mu_x}(y)$ refer to the $y$ and $x$ values attaining the optima
in~\eqref{eq:smoothed_functions}. These can be thought of as
\emph{smoothed best responses}. \citet{Nesterov05:Smooth} shows that the
gradients of the functions $f_{\mu_y}(x)$ and $\phi_{\mu_x}(y)$ exist and are
Lipschitz continuous. The gradient operators and Lipschitz constants are 
\begin{align*}
  &\nabla f_{\mu_y}(x) = a_1 + Ay_{\mu_y}(x),  &\nabla \phi_{\mu_x}(y) = a_2 + A^\top x_{\mu_x}(y),\\
  &L_1\left(f_{\mu_y}\right) = \frac{\|A\|^2}{\varphi_\cY\mu_y}, &L_2\left(\phi_{\mu_x}\right) = \frac{\|A\|^2}{\varphi_\cX\mu_x},
\end{align*}
where $\|A\|$ is the $\ell_1$-norm operator norm.

Let the convex conjugate of $d_{\cX}:\cX \rightarrow \bbR$ be denoted by
$d_{\cX}^*(g)=\max_{x\in \cX}g^Tx - d(x)$. The gradient $\nabla d^*(g)$ of the
conjugate then gives the solution to the smoothed-best-response problem.

Based on this setup, {\egt} minimizes the following saddle-point residual, which
is equal to the sum of regrets for the players.
\begin{align*}
 \epsilonsad(x^t,y^t) = \max_{y\in\cY}(x^t)^TAy-\min_{x\in\cX}x^TAy^t
\end{align*}
The idea behind EGT is to maintain the \emph{excessive gap condition} (EGC),
% \begin{align*}
  $
 \textsc{EGV}(x,y) \defeq   \phi_{\mu_x}(y) - f_{\mu_y}(x) > 0.
 $
% \end{align*}
The EGC implies a bound on the saddle-point residual: $ \epsilonsad(x^t,y^t)
\leq \mu_x \Omega_\cX + \mu_y\Omega_\cY$, where $\Omega_\cX = \max_{x,x'} d_\cX(x) - d_\cX(x')$, and $\Omega_\cY$ defined analogously.

We formally state {\egt}~\cite{Nesterov05:Excessive} as
Algorithm~1.
\begin{figure}
\begin{minipage}{.52\linewidth}
  \vspace{-4mm}
\begin{algorithm}[H]\small
 \caption{\textsc{EGT}(DGF-center $x_\omega$, DGF weights $\mu_x,\mu_y$, and $\epsilon>0$)}
 \label{alg:egt}
\begin{algorithmic}[1]
% \PROCEDURE{EGT}{$\omega$-center $z_\omega$, DGF weights $\mu_x,\mu_y$, and $\epsilon>0$}
\STATE $x^{0} = \nabla d_{\cX}^*\left( \mu_x^{-1} \nabla
  f_{\mu_y}(x_{\omega})\right)$
\STATE $y^0 = y_{\mu_y}(x_{\omega})$
\STATE $t = 0$
\WHILE{$\epsilonsad(x^t,y^t)>\epsilon$}
  \STATE $\tau_t = \frac{2}{t+3}$
  \IF{$t$ is even}
    \STATE $(\mu_x^{t+1},x^{t+1},y^{t+1}) = \textsc{Step}(\mu_x^{t}, \mu_y^t, x^t, y^t, \tau)$
  \ELSE
    \STATE$(\mu_y^{t+1},y^{t+1},x^{t+1}) = \textsc{Step}(\mu_y^t, \mu_x^t, y^t, x^t, \tau)$
  \ENDIF
  \STATE $t=t+1$
\ENDWHILE
\STATE \textbf{return} $x^t,y^t$
% \ENDPROCEDURE
\end{algorithmic}
\end{algorithm}
\end{minipage}
\begin{minipage}{.48\linewidth}
\begin{algorithm}[H]\small
  \caption{\textsc{Step}($\mu_x,\mu_y, x, y, \tau$)}
  \label{alg:step}
% \SetKwInOut{Inp}{input}
% \SetKwInOut{Output}{output}
\begin{algorithmic}[1]
% \Inp{$\mu_x,\mu_y, x, y, \tau$}
  \STATE $\hat{x} = \left( 1-\tau \right)x + \tau x_{\mu_x}(y)$
  \STATE $y_+ = \left( 1-\tau \right)y + \tau y_{\mu_y}(\hat{x})$
  \STATE $\tilde{x} = \nabla d_\cX^*{}\left(\nabla d_{\cX}(x_{\mu_x}(y)) - \frac{\tau}{\left( 1-\tau \right)\mu_x} \nabla f_{\mu_y}(\hat{x})\right)$
  \STATE $x_+ = \left( 1-\tau \right)x + \tau \tilde{x}$
  \STATE $\mu_x^+ = \left( 1-\tau \right)\mu_x$
  \STATE \textbf{return} $\mu_x^+,x_+,y_+$
\end{algorithmic}
\end{algorithm}
\end{minipage}
\end{figure}
% \begin{figure}
% \centering
%    \includegraphics[width=.9\linewidth]{../figs/egt1.pdf}
%    \includegraphics[width=.9\linewidth]{../figs/egt2.pdf}
% \end{figure}
The \egt\ algorithm alternates between taking steps focused on $\cX$ and $\cY$. Algorithm~2 shows a single step focused on $\cX$. Steps focused on $y$ are analogous. Algorithm~1 shows how the alternating steps and stepsizes are computed, as well as how initial points are selected.

\noindent Suppose the initial values $\mu_x,\mu_y$ satisfy
$\mu_x=\frac{\varphi_\cX}{L_1(f_{\mu_y})}$. Then, at every iteration $t\geq 1$
of {\egt}, the corresponding solution $z^t=[x^t;y^t]$ satisfies $x^t\in \cX$,
$y^t\in \cY$, the excessive gap condition is maintained, and
\[
\vspace{-1mm}
\epsilonsad(x^T,y^T) \leq  \frac{4\|A\|}{T+1}\sqrt{\frac{\Omega_\cX\Omega_\cY}{\varphi_\cX\varphi_\cY}}.
% f(x^t)-\phi(y^t) = \epsilonsad(z^t) \leq  \frac{4\|A\|}{T+1}\sqrt{\frac{\Omega_\cX\Omega_\cY}{\varphi_\cX\varphi_\cY}}.
\vspace{-1mm}
\]
Consequently, {\egt} has a convergence rate of $O(\frac{1}{T})$~\cite{Nesterov05:Excessive}.
%%% Local Variables:
%%% mode: latex
%%% TeX-master: "../ijcai18/gpu_egt_river_ijcai18"
%%% End:

% DGFs for EFGs
\section{Treeplexes}
\label{sec:treeplexes}
\citet{Hoda10:Smoothing} introduced the {\em treeplex}, a class of convex polytopes that captures the sequence-form of the strategy spaces in perfect-recall {\efg}s.
%We use the definition from \citet{Kroer15:Faster}:
\begin{definition}
  Treeplexes are defined recursively:
  \begin{enumerate}
  \item {\em Basic sets\/}: The standard simplex $\Delta_m$ %$\Delta_m = \left\{ x \in \left[0,1\right]^m  : \sum_{k=1}^m x_k = 1 \right\}$
  is a treeplex.
  \item {\em Cartesian product\/}:  If $Q_1,\ldots, Q_k$ are treeplexes, then $Q_1 \times \cdots \times Q_k$ is a treeplex.
  \item {\em Branching\/}: Given a treeplex $P\subseteq \left[ 0,1 \right]^p$, a collection of treeplexes $Q=\left\{ Q_1,\ldots,Q_k \right\}$ where $Q_j\subseteq \left[ 0,1 \right]^{n_j}$, and $l=\left\{l_1,\ldots,l_k \right\} \subseteq \left\{ 1,\ldots, p \right\}$, the set defined by
\begin{align*}
\vspace{-1mm}
  P\framebox{l}Q \coloneqq \left\{ \left(x,y_1,\ldots,y_k\right) \in \R^{p+\sum_j n_j}  :~ x\in P,
    \, y_1\in x_{l_1} \cdot Q_1,\, \ldots, y_k\in x_{l_k} \cdot Q_k \vphantom{\R^{\sum}}\right\}
\vspace{-1mm}
\end{align*}
%    \begin{align*}
    %& P\framebox{l}Q \coloneqq \left\{ \left(x,y_1,\ldots,y_k\right) \in \R^{p+\sum_j q_j} \right. \\
      %& \left. :~ x\in P,\, y_1\in x_{l_1} \cdot Q_1,\, \ldots, y_k\in x_{l_k} \cdot Q_k \right\}
%    \end{align*}
    is a treeplex.
    We say $x_{l_j}$ is the branching variable for the treeplex $Q_j$.
  \end{enumerate}
\end{definition}
One interpretation of the treeplex is as a set of simplexes, where each simplex
is weighted by the value of the variable above it in the parent branching
operation (or $1$ if there is no branching operation preceding the simplex).
Thus the simplexes generally sum to the value of the parent rather than $1$.

For a treeplex $Q$, we denote by $S_Q$ the index set of the set of simplexes
contained in $Q$ (in an \efg\ $S_Q$ is the set of information sets belonging to
the player). For each $j\in S_Q$, the treeplex rooted at the $j$-th simplex
$\Delta^j$ is referred to as $Q_j$. Given vector $q\in Q$ and simplex
$\Delta^j$, we let $\bbI_j$ denote the set of indices of $q$ that correspond to
the variables in $\Delta^j$ and define $q^j$ to be the subvector of $q$
corresponding to the variables in $\bbI_j$. For each simplex $\Delta^j$ and
branch $i\in \bbI_j$, the set $\cD_j^i$ represents the set of indices of
simplexes reached immediately after $\Delta^j$ by taking branch $i$ (in an \efg,
$\cD_j^i$ is the set of potential next-step information sets for the player).
Given a vector $q\in Q$, simplex $\Delta^j$, and index $i\in \bbI_j$, each child
simplex $\Delta^k$ for every $k\in \cD_j^i$ is scaled by $q_i$. For a given
simplex $\Delta^j$, we let $p_j$ denote the index in $q$ of the parent branching
variable $q_{p_j}$ scaling $\Delta^j$. We use the convention that $q_{p_j}=1$ if
$Q$ is such that no branching operation precedes $\Delta^j$. For each $j\in
S_Q$, $d_j$ is the maximum depth of the treeplex rooted at $\Delta^j$, that is,
the maximum number of simplexes reachable through a series of branching
operations at $\Delta^j$. Then $d_Q$ gives the depth of $Q$. We use $b_Q^j$ to
identify the number of branching operations preceding the $j$-th simplex in $Q$.
We say that a simplex $j$ such that $b_Q^j=0$ is a \emph{root simplex}.

\begin{figure}[!h]
  \begin{center}
    \includegraphics[width=0.9\textwidth]{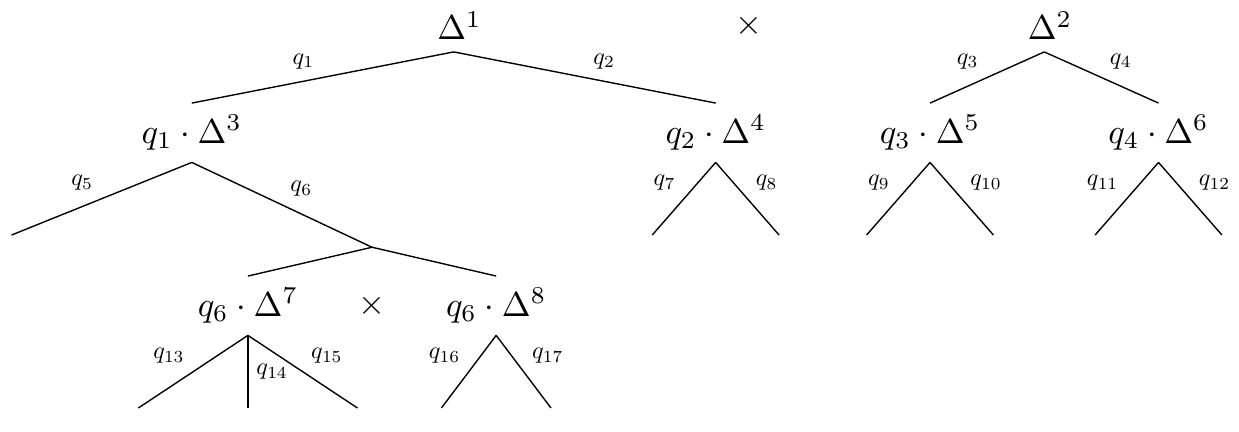}
  \end{center}
  \caption{An example treeplex constructed from $9$ simplexes. Cartesian product operation is denoted by $\times$.  }
  \label{fig:treeplex}
\end{figure}

Figure~\ref{fig:treeplex} illustrates an example treeplex $Q$. This treeplex $Q$ is
constructed from nine two-to-three-dimensional simplexes $\Delta^1,\ldots,
\Delta^9$. At level~$1$, we have two root simplexes, $\Delta^1,\Delta^2$,
obtained by a Cartesian product operation (denoted by $\times$). We have maximum depths $d_1=2$, $d_2=1$ beneath them.
Since there are no preceding branching operations, the parent variables for
these simplexes $\Delta^1$ and $\Delta^2$ are $q_{p_1}=q_{p_2}=1$. For
$\Delta^1$, the corresponding set of indices in the vector $q$ is
$\bbI_1=\left\{ 1,2 \right\}$, while for $\Delta^2$ we have $\bbI_2=\left\{ 3,4,
  5 \right\}$. At level~$2$, we have the simplexes $\Delta^3,\ldots,\Delta^7$.
The parent variable of $\Delta^3$ is $q_{p_3}=q_1$; therefore, $\Delta^3$ is
scaled by the parent variable $q_{p_3}$. Similarly, each of the simplexes
$\Delta^3,\ldots,\Delta^7$ is scaled by their parent variables $q_{p_j}$ that
the branching operation was performed on. So on for $\Delta^8$ and $\Delta^9$ as
well. The number of branching operations required to reach simplexes
$\Delta^1,\Delta^3$ and $\Delta^8$ is $b_Q^1=0,b_Q^3=1$ and $b_Q^8=2$,
respectively.

%%% Local Variables:
%%% mode: latex
%%% TeX-master: "nips18/gpu_egt_river_nips18"
%%% End:

\section{Smoothed Best Responses}

Let $d_j(x) = \sum_{i \in \bbI_j} x_i\log{x_i} + \log{n}$ be the entropy DGF for the $n$-dimensional simplex $\Delta_n$, where
$n$ is the dimension of the $j$'th simplex in $Q$. \citet{Kroer17:Theoreticala}
introduced the following DGF for $Q$ by \emph{dilating} $d_s$ for each simplex in
$S_Q$ and take their sum:
% \begin{equation}
  $
  \label{eq:dilated_entropy}
    d(q) = \sum_{j\in S_Q} \beta_j q_{p_j}d_j\left(\frac{q^j}{q_{p_j}} \right),
    $
% \end{equation}
where $\beta_j = 2 + \sum_{k \in \cD^j} 2 \beta_k$. Other dilated DGFs for
treeplexes were introduced by \citet{Hoda10:Smoothing} and were 
also studied by \citet{Kroer15:Faster}. \citet{Kroer17:Theoreticala} proved that
this DGF is strongly convex modulus $\frac{1}{M}$ where $M$ is the maximum value
of the $\ell_1$ norm over $Q$. EGT instantiated with this DGF converges at a
rate of $\frac{LM^22^{d}\log n}{T}$ where $L$ is the maximum entry in the
payoff matrix, $d$ is the depth of the treeplex, and $n$ is the maximum
dimension of any individual simplex.

We now show how to solve~\eqref{eq:smoothed_functions} for this particular DGF\@. While it
is known that this DGF has a closed-form solution, this is the first time the
approach has been shown in a paper. Furthermore, we believe that our particular solution
is novel, and leads to better control over numerical issues.
The problem we wish to solve is the following.
\begin{align}
  \argmin \sum_{j \in S_Q} \langle q^j, g_j \rangle + \beta_j q_{p_j} d_j(q^j / q_{p_j})
  = \argmin \sum_{j \in S_Q} q_{p_j} (\langle \bar{q}^j , g_j\rangle + \beta_j d_j(\bar{q}^j)) \label{eq:dilated_entropy_sum}
\end{align}
where the equality follows by the fact that $q_i=q_{p_j}\bar{q}_i$. For a leaf
simplex $j$, its corresponding term in the summation has no dependence on any
other part of the game tree except for the multiplication by $x_{p_j}$ (because none of its variables are parent to any other
simplex). Because of this lack of dependence, the expression
\[
 \langle \bar{q}^j / q_{p_j}, g_j\rangle + \beta_j d_j(q^j / q_{p_j})
\]
can be minimized independently as if it were an optimization problem over a
simplex with variables $\bar{q}^j=x^j/q_{p_j}$ (this was also pointed out in
Proposition 3.4 in \citet{Hoda10:Smoothing}). We show how to solve the
optimization problem at a leaf:
% \begin{align*}
  $
 \min_{\bar{q}^j \in \Delta_{j}}\ \langle \bar{q}^j, g_j\rangle + \beta_j d_j(\bar{q}^j).
 $
% \end{align*}
Writing the Lagrangian with respect to the simplex constraint and taking the
derivative wrt. $\bar{q}_i$ gives
\begin{align*}
 \min_{\bar{q}^j}\ \langle \bar{q}^j, g_j\rangle + \beta_j d_j(\bar{q}^j) + \lambda (1 - \sum_{i\in \bbI_j} \bar{q}_i) \Rightarrow
% \end{align*}
% Taking the derivative wrt. $\bar{q}_i$ gives the well-known solution
% \begin{align*}
  g_i + \beta_j (1+\log{\bar{q}_i}) = \lambda \Rightarrow \bar{q}_i \propto e^{-g_i / \beta_j}
\end{align*}
This shows how to solve the smoothed-best-response problem at a leaf. For an
internal simplex $j$, Proposition 3.4 of \citet{Hoda10:Smoothing} says that we
can simply compute the value at all simplexes below $j$, add the value to $g_j$
(this is easily seen from~\eqref{eq:dilated_entropy_sum}; each $q_i$ acts as a scalar on the value of all simplexes after $i$), and proceed by
induction.
Letting $|\bbI_j|=n$, we now simplify the objective function:
\begin{align*}
  & \langle \bar{q}^j, g_j \rangle + \beta_j (\sum_{i\in \bbI_j} (\bar{q}_{i} \log \bar{q}_{i}) + \log n)
  = \sum_i (\bar{q}_{i}  (g_{i} + \beta_j \log \bar{q}_{i})) + \beta_j \log n \\
  =& \sum_i (\bar{q}_{i}  (\lambda - \beta_j)) + \beta_j \log n
  = \lambda - \beta_j + \beta_j \log n,
\end{align*}
where the last two equalities follow first by applying our derivation for
$\lambda$ and then the fact that $\bar{q}^j$ sums to one.
This shows that we can choose an arbitrary index $i\in \bbI_j$ and propagate the
value $ g_{i} + \beta_j \log \bar{q}_{i} + \beta_j \log n $. In particular, for
numerical reasons we choose the one that maximizes $\bar{q}_i$.

In addition to smoothed best responses, fast FOMs usually also require
computation of proximal mappings, which are solutions to 
% the following
% optimization problem
% \begin{align*}
  $
  \argmin_{q\in Q}\ \langle q, g \rangle + D(q \| q'),
  $
% \end{align*}
where $D(q \| q')= d(q) - d(q') - \langle \nabla d(q'), q-q' \rangle$ is the
Bregman divergence associated with the chosen DGF $d$. Unlike the smoothed best
response, we are usually only interested in the minimizing solution and not the
associated value. Therefore we can drop terms that do not depend on $q$ and the
problem reduces to
% \begin{align*}
  $
  \argmin_{q\in Q}\ \langle q, g \rangle + d(q) - \langle \nabla d(q'), q \rangle,
  $
% \end{align*}
which can be solved with our smoothed best response approach by using the
shifted gradient $\tilde{g} = g - \nabla d(q')$. This has one potential
numerical pitfall: the DGF-gradient $\nabla d(q')$ may be unstable near the
boundary of $Q$, for example because the entropy DGF-gradient requires taking
logarithms. It is possible to derive a separate expression for the proximal
mapping that is similar to what we did for the smoothed best response; this
expression can help avoid this issue. However, because we only care about
getting the optimal solution, not the value associated with it, this is not
necessary. The large gradients near the boundary only affect the solution by
setting bad actions too close to zero, which does not seem to affect
performance.

%%% Local Variables:
%%% mode: latex
%%% TeX-master: "../nips1818/gpu_egt_river_nips1818"
%%% End:

% practical EGT 
\section{Practical EGT}
Rather than the overly conservative stepsize and $\mu$ parameters suggested in
the theory for EGT we use more practical variants combining practical techniques
from \citet{Kroer17:Theoreticala} and \citet{Hoda10:Smoothing}. The pseudocode
is shown in Algorithm~\ref{alg:aggressive_egt}. As in
\citet{Kroer17:Theoreticala} we use a practically-tuned initial choice for the
initial smoothing parameters $\mu$. Furthermore, rather than alternating the
steps on players 1 and 2, we always call \textsc{Step} on the player with a
higher $\mu$ value (this choice is somewhat reminiscent of the $\mu$-balancing
heuristic employed by \citet{Hoda10:Smoothing} although our approach avoids an
additional fitting step). The EGT algorithm with a practically-tuned $\mu$ and
this $\mu$ balancing heuristic will be denoted EGT in our experiments. In
addition, we use an EGT variant that employs the \emph{aggressive $\mu$ reduction}
technique introduced by \citet{Hoda10:Smoothing}. Aggressive $\mu$ reduction uses
the observation that the original EGT stepsize choices, which are
$\tau=\frac{2}{3+t}$, are chosen to guarantee the excessive gap condition, but
may be overly conservative. Instead, aggressive $\mu$ reduction simply maintains
some current $\tau$, initially set to $0.5$, and tries to apply the same
stepsize $\tau$ repeatedly. After every step, we check that the excessive gap
condition still holds; if it does not hold then we backtrack, $\tau$ is
decreased, and we repeat the process. A $\tau$ that maintains the condition is
always guaranteed to exist by Theorem 2 of \citet{Nesterov05:Excessive}. The
pseudocode for this is given in Algorithm~\ref{alg:aggressive_step}. EGT with
aggressive $\mu$ reduction, a practically tuned initial $\mu$, and $\mu$ balancing,
will be denoted \textsc{EGT/as} in our experiments.

\begin{figure}
\begin{minipage}{.58\linewidth}
  \vspace{-4mm}
\begin{algorithm}[H]\small
	\caption{\textsc{EGT/as}(DGF-center $x_\omega$, DGF weights $\mu_x,\mu_y$, and $\epsilon>0$)}
	\label{alg:aggressive_egt}
	\begin{algorithmic}[1]
		% \PROCEDURE{EGT}{$\omega$-center $z_\omega$, DGF weights $\mu_x,\mu_y$, and $\epsilon>0$}
		\STATE $x^{0} = \nabla d_{\cX}^*\left( \mu_x^{-1} \nabla
		f_{\mu_y}(x_{\omega})\right)$
		\STATE $y^0 = y_{\mu_y}(x_{\omega})$
		\STATE $t = 0$
		\STATE $\tau = \frac{1}{2}$
		\WHILE{$\epsilonsad(x^t,y^t)>\epsilon$}
		\IF{$\mu_x > \mu_y$}
		\STATE $(\mu_x^{t+1},x^{t+1},y^{t+1},\tau) = \textsc{Decr}(\mu_x^{t}, \mu_y^t, x^t, y^t, \tau)$
		\ELSE
		\STATE$(\mu_y^{t+1},y^{t+1},x^{t+1},\tau) = \textsc{Decr}(\mu_y^t, \mu_x^t, y^t, x^t, \tau)$
		\ENDIF
		\STATE $t=t+1$
		\ENDWHILE
		\STATE \textbf{return} $x^t,y^t$
		% \ENDPROCEDURE
	\end{algorithmic}
\end{algorithm}
\end{minipage}
\begin{minipage}{.42\linewidth}
\begin{algorithm}[H]\small
	\caption{\textsc{Decr}($\mu_x,\mu_y, x, y, \tau$)}
\label{alg:aggressive_step}
	\begin{algorithmic}[1]
		% \PROCEDURE{EGT}{$\omega$-center $z_\omega$, DGF weights $\mu_x,\mu_y$, and $\epsilon>0$}
		\STATE $(\mu_x^+,x^+,y^+) = \textsc{Step}(\mu_x, \mu_y, x, y, \tau)$
		\WHILE{\textsc{EGV}$(x,y) < 0$}
		\STATE $\tau = \frac{1}{2}\tau$
		\STATE $(\mu_x^{+},x^{+},y^{+}) = \textsc{Step}(\mu_x, \mu_y, x, y, \tau)$
		\ENDWHILE
		\STATE \textbf{return} $\mu_x^{+}x^t,y^t,\tau$
		% \ENDPROCEDURE
	\end{algorithmic}
\end{algorithm}
  \vspace{-4mm}
\end{minipage}
\end{figure}
%%% Local Variables:
%%% mode: latex
%%% TeX-master: "../nips1818/gpu_egt_river_nips1818"
%%% End:

% CFR
% \input{text/cfr}
% GPU implementation
\section{Algorithm Implementation}

To compute smoothed best responses, we use a parallelization scheme. 
% that works
%for any treeplex where the top of the treeplex consists of a wide Cartesian
%product. 
We parallelize across the initial Cartesian product of treeplexes at the root.
As long as this Cartesian product is wide enough, the smoothed best response
computation will take full advantage of parallelization. This is a common
structure in real-world problems, for example representing the starting hand in
poker, or some stochastic private state of each player in other applications.
This parallelization scheme also works for gradient computation based on tree
traversal. However, in this paper we do gradient computation by writing down a
sparse payoff matrix using CUDA's sparse library and let CUDA parallelize the
gradient computation.

For poker-specific applications (and certain other games where utilities
decompose nicely based on private information) it is possible to speed up the
gradient computation substantially by employing the accelerated tree traversal
of \citet{Johanson11:Accelerating}. We did not use this technique. In our
experiments, the majority of time is spent in gradient computation, so this
acceleration is likely to affect all the tested algorithms equally. Furthermore, since
the technique is specific to games with certain structures, our experiments
give a better estimate of general EFG-solving performance.

%%% Local Variables:
%%% mode: latex
%%% TeX-master: "../ijcai18/gpu_egt_river_ijcai18"
%%% End:

% Experiments

\section{Experiments}

We now present experimental results on running all the previously described
algorithms on a GPU\@. All experiments were run on a Google Cloud instance with
an NVIDIA Tesla K80 GPU with 12GB available. All code was implemented in C++
using CUDA for GPU operations, and cuSPARSE for the sparse payoff matrix.
We compare against several CFR variants.\footnote{All variants use the
	alternating updates scheme.} We run CFR with RM (CFR(RM)), {\rmp} (CFR({\rmp})),
and {\cfrp} which is CFR with {\rmp} and a linear averaging scheme. We now
describe these variants. Detailed descriptions can also be found in
\citet{Zinkevich07:Regret} and \citet{Tammelin15:Solving}.

Our experiments are conducted on real large-scale ``river'' endgames faced by
the \emph{Libratus} AI~\citep{Brown17:Superhuman}. \emph{Libratus} was created
for the game of heads-up no-limit Texas hold'em.
% (NLTH).
% In NLTH
% one player, the big blind, puts \$100 in the pot initially, and the other
% player, the small blind, puts \$50 in the pot (in our experiments Libratus is
% the big blind).
% Afterwards, Chance deals out a pair of cards (a hand) to each
% player from a standard 52-card deck. Then the first betting round occurs. After
% the betting round, if neither player folded then Chance deals out
%
\emph{Libratus} was constructed by first computing a ``blueprint'' strategy for
the whole game (based on abstraction and Monte-Carlo
CFR~\citep{Lanctot09:Monte}). Then, during play, \emph{Libratus} would solve
endgames that are reached using a significantly finer-grained abstraction. In
particular, those endgames have no card abstraction, and they have a
fine-grained betting abstraction.
%The publicly observed information is held constant
%(i.e. bets made and board cards), while the players' hands are stochastic.
For the beginning of the subgame, the blueprint strategy gives a conditional
distribution over hands for each player. The subgame is constructed by having a
Chance node deal out hands according to this conditional
distribution.\footnote{\emph{Libratus} used two different subgame-solving
	techniques, one ``unsafe'' and one ``safe''~\citep{Brown17:Safe}. The
	computational problem in the two is essentially identical. We experiment with
the ``unsafe'' version, which uses the prior distributions described here.}

A subgame is structured and parameterized as follows. The game is
parameterized by the conditional distribution over hands for each player,
current pot size, board state (5 cards dealt to the board), and a betting
abstraction. First, Chance deals out hands to the two players according to the
conditional hand distribution. Then, Libratus has the choice of folding,
checking, or betting by a number of multipliers of the pot size: 0.25x, 0.5x,
1x, 2x, 4x, 8x, and all-in. If Libratus checks and the other player bets then
Libratus has the choice of folding, calling (i.e.\ matching the bet and ending
the betting), or raising by pot multipliers 0.4x, 0.7x, 1.1x, 2x, and all-in. If
Libratus bets and the other player raises Libratus can fold, call, or raise by
0.4x, 0.7x, 2x, and all-in. Finally when facing subsequent raises Libratus can
fold, call, or raise by 0.7x and all-in. When faced with an initial check, the
opponent can fold, check, or raise by 0.5x, 0.75x, 1x, and all-in. When faced
with an initial bet the opponent can fold, call, or raise by 0.7x, 1.1x, and
all-in. When faced with subsequent raises the opponent can fold, call, or raise
by 0.7x and all-in. The game ends whenever a player folds (the other player wins
all money in the pot), calls (a showdown occurs), or both players check as their
first action of the game (a showdown occurs). In a showdown the player with the 
better hands wins the pot. The pot is split in case of a tie. (For our experiments we used endgames where it is \emph{Libratus}'s turn to move first.)
% in our
%experiments Libratus always goes first, i.e. is the big blind.
%For the case
%where Libratus goes second the betting abstraction would change slightly, but we
%expect the results to be very similar.

We conducted experiments on two river endgames extracted from \emph{Libratus}
play: Endgame 2 and Endgame 7. Endgame 2 has a pot of size $2100$ at the
beginning of the river endgame. It has dimension 140k and 144k for
\emph{Libratus} and the opponent, respectively, and 176M leaves in the games
tree. Endgame 7 has a pot of size \$3750 at the beginning of the river subgame.
It has dimension 43k and 86k for the players, and 54M leaves.

\begin{figure}[t]
	\includegraphics[width=0.49\columnwidth]{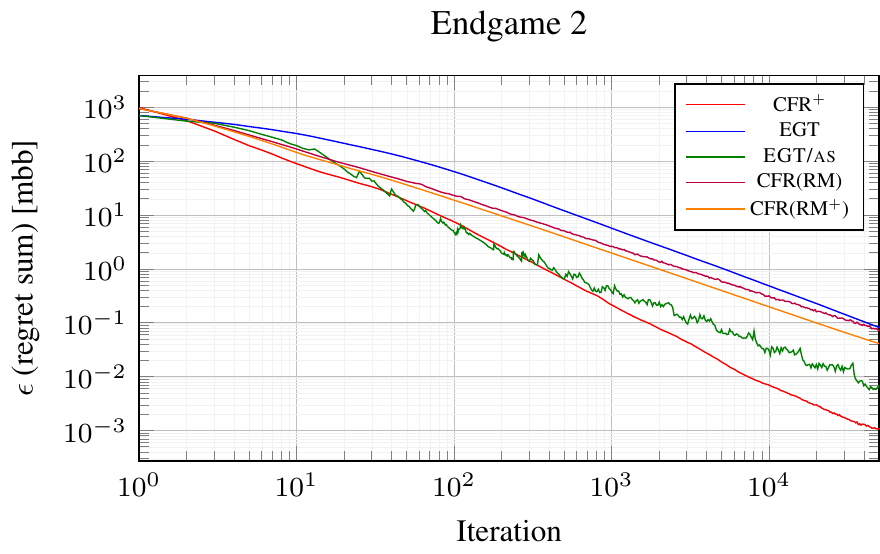}
	\includegraphics[width=0.49\columnwidth]{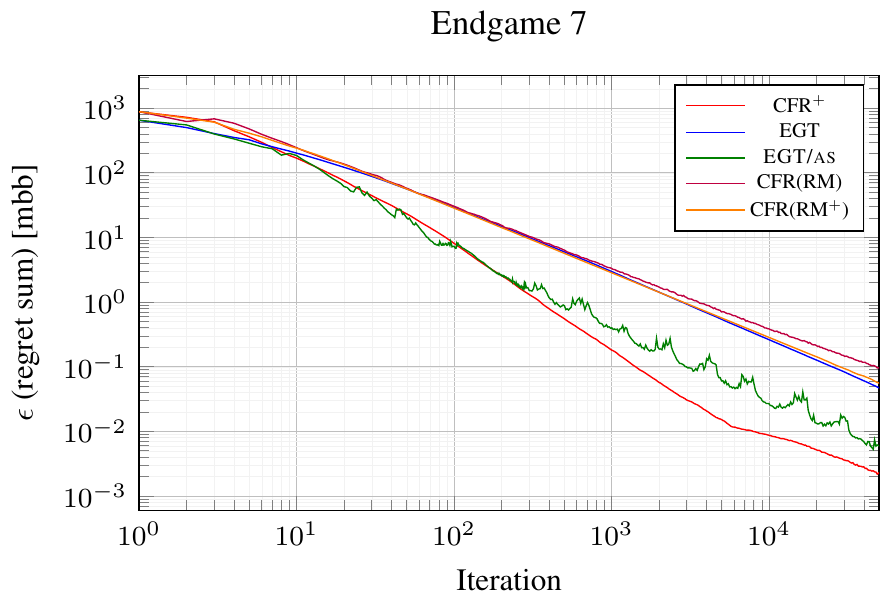}
	\caption{Solution quality as a function of the number of iterations for all
		algorithms on two river subgames. The solution quality is given as the sum
	of regrets for the players in milli-big-blinds.}
	\label{fig:per_iteration}
\end{figure}
In the first set of experiments we look at the per-iteration performance of each
algorithm. The results are shown in Figure~\ref{fig:per_iteration}. The y-axis
shows the sum of the regrets for each player, that is, how much utility they can
gain by playing a best response rather than their current strategy. The unit is
milli-big-blinds (mbb); at the beginning of the original poker game,
\emph{Libratus}, as the ``big blind'', put in \$100 and the opponent put in
\$50, in order to induce betting. Mbb is a thousandth of the big blind value,
that is, 10 cents. This is a standard unit used in research that uses poker
games for evaluation. One mbb is often considered the convergence goal. {\cfrp}
and \textsc{EGT/as} perform the best; both reach the goal of 1mbb after about
400 iterations in both Endgame~2 and~7. EGT, CFR(RM), and CFR(\rmp) all take
about 3000 iterations to reach 1mbb in Endgame 7. In Endgame~2, EGT is slowest,
although the slope is steeper than for CFR(RM) and CFR(\rmp). We suspect that
better initialization of EGT could lead to it beating both algorithms. Note also
that EGT was shown better than CFR(RM) and CFR(\rmp) by
\citet{Kroer17:Theoreticala} in the smaller game of Leduc hold'em with an
automated $\mu$-tuning approach. Their results further suggest that better
initialization may help enhance converge speed significantly.

\begin{figure}[t]
	\includegraphics[width=0.49\columnwidth]{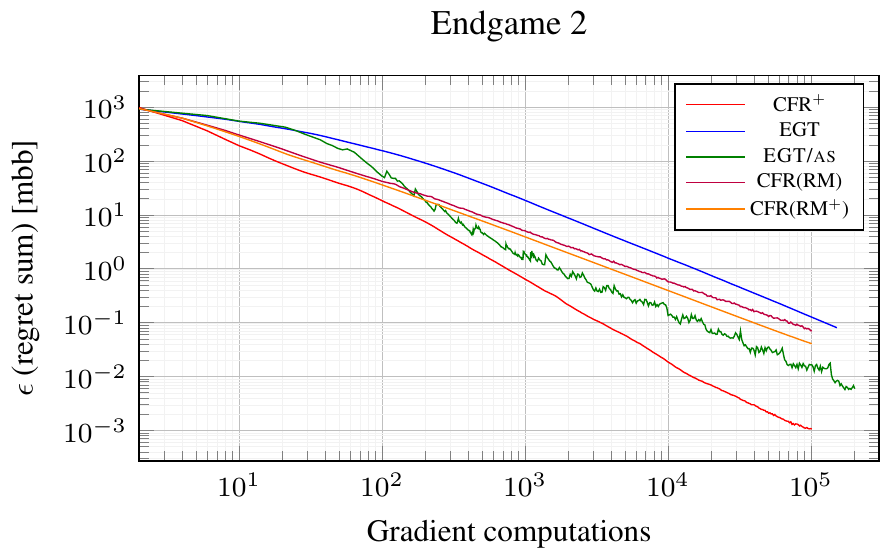}
	\includegraphics[width=0.49\columnwidth]{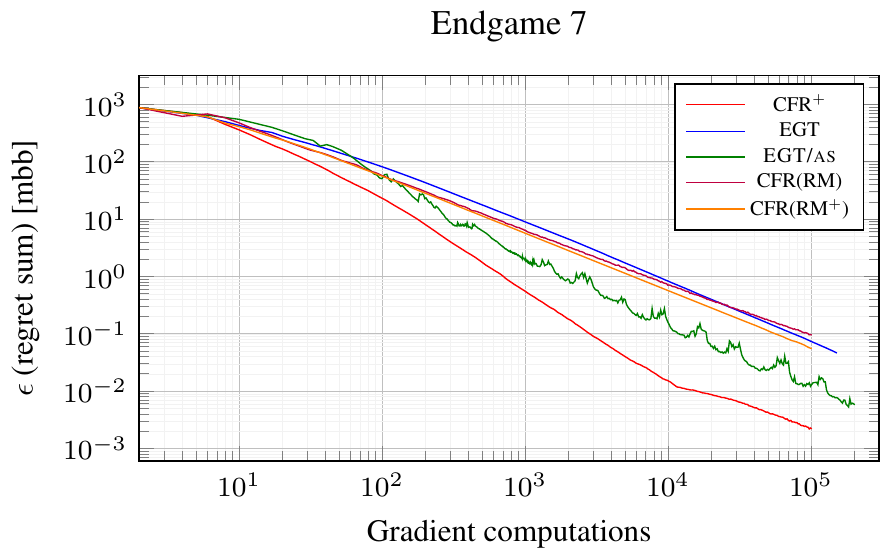}
	\caption{Solution quality as a function of the number of gradient
		computations for all algorithms on two river subgames. The solution quality
	is given as the sum of regrets for the players in milli-big-blinds.}
	\label{fig:per_gradient}
	\vspace{-2mm}
\end{figure}

One issue with per-iteration convergence rates is that the algorithms do not
perform the same amount of work per iteration. All CFR variants in our
experiments compute 2 gradients per iteration, whereas EGT computes 3, and
\textsc{EGT/as} computes 4 (the additional gradient computation is needed in
order to evaluate the excessive gap). Furthermore, \textsc{EGT/as} may use
additional gradient computations if the excessive gap check fails and a smaller
$\tau$ is tried (in our experiments about 15 adjustments were needed). In our
second set of plots, we show the convergence rate as a function of the total
number of gradient computations performed by the algorithm. This is shown in
Figure~\ref{fig:per_gradient}. By this measure, \textsc{EGT/as} and EGT perform
slightly worse relative to their performance as measured by iteration count. In
particular, {\cfrp} takes about 800 gradient computations in order to reach 1mbb
in either game, whereas \textsc{EGT/as} takes about 1800.

In our experiments {\cfrp} vastly outperforms its theoretical convergence rate
(in fact, every CFR variant does significantly better than the theory predicts,
but {\cfrp} especially so). However, {\cfrp} is known to eventually reach a
point where it slows down and performs worse than $\frac{1}{T}$. In our
experiments we start to see {\cfrp} slowing down towards the end of Endgame 7.
EGT, in contrast, is guaranteed to maintain a rate of $\frac{1}{T}$, and so may
be preferable if a guarantee against slowdown is desired or high precision is
needed.

%%% Local Variables:
%%% mode: latex
%%% TeX-master: "nips2018/gpu_egt_river_nips18"
%%% End:

% Conclusion
\section{Conclusions and Future Research}

We introduced a practical variant of the EGT algorithm that employs aggressive
stepsizes, $\mu$ balancing, a numerically-friendly smoothed-best-response
algorithm, parallelization via Cartesian product operations at the root of the
strategy treeplex, and a GPU implementation. We showed for the first time, via
experiments on real large-scale \emph{Libratus} endgames, that FOMs (with the
dilated entropy DGF) are competitive with the CFR family of algorithms.
Specifically, they outperform the other CFR variants and are close in efficiency
to {\cfrp}. Our best variant of EGT can solve subgames to the desired accuracy
at a speed that is within a factor of two of {\cfrp}. 

Our results suggest that it may be possible to make FOMs faster than {\cfrp}.
For example, we did not spend much effort tuning the parameters of EGT, and
tuning them would make the algorithm even more efficient. Second, we only
investigated EGT, which has been most popular FOM in EFG solving. However, it is
possible that other FOMs such as mirror prox~\citep{Nemirovski04:Prox} or the
primal-dual algorithm by \citet{Chambolle11:First} could be made even faster. 

Furthermore, stochastic FOMs (i.e., ones where the gradient is approximated by
sampling to make the gradient computation dramatically faster) could be
investigated as well. \citet{Kroer15:Faster} tried this using \emph{stochastic
mirror prox}~\citep{Juditsky11:Solving} without practical success, but it is
likely that this approach could be made better with more engineering. 

It would also be interesting to compare our EGT approach to CFR algorithms for
computing equilibrium refinements, for example in the approximate extensive-form
perfect equilibrium model investigated by \citet{Kroer17:Smoothing} and
\citet{Farina17:Regret}.

Pruning techniques (for temporarily skipping parts of the game tree on some
iterations) have been shown effective for both CFR and EGT-like algorithms, and
could potentially be incorporated as
well~\citep{Lanctot09:Monte,Brown17:Dynamic,Brown17:Reduced}.

Finally, while EGT, as well as other FOM-based approaches to computing zero-sum
Nash equilibria, are not applicable to the computation of general-sum Nash
equilibria in theory they could still be applied to the computation of
strategies in practice (gradients can still be computed, and so the smoothed
best responses and corresponding strategy updates are still well-defined). For
CFR the analogous approach seems to perform reasonably
well~\citep{Cermak15:Strategy}, and you might expect the same from FOMs such as
EGT.

%%% Local Variables:
%%% mode: latex
%%% TeX-master: "../nips18/gpu_egt_river_nips18"
%%% End:

\noindent {\bf Acknowledgments}
This material is based on work supported by the National
Science Foundation under grants IIS-1718457, IIS-1617590,
and CCF-1733556, and the ARO under award W911NF-17-1-
0082. Christian Kroer is supported by a Facebook Fellowship.

\bibliographystyle{abbrvnat}
\bibliography{dairefs}

\newpage
\appendix

\section{Prox shift}

We want to find a simple, numerically-friendly expression for
\begin{align*}
	-d(q) + \langle \nabla d(q), q \rangle 
\end{align*}

First we derive an expression for $\langle \nabla d(q), q \rangle$. Let $i\in
\bbI_j$ and $n_j$ be the dimensionality of simplex $j$. Taking derivatives we
have
\begin{align*}
	\nabla_{ji} d(q) 
	& = \beta_j( \log \frac{q_{i}}{q_{p_j}} + 1) 
	+ \sum_{k \in \cD_j^i} \beta_k \left( 
		\log n_k - \sum_{i'\in \bbI_k} \frac{x_{i'}}{x_{p_k}}
	\right) \\
	& = \beta_j(\log \frac{q_{i}}{q_{p_j}} + 1) 
	+ \sum_{k \in \cD_j^i} \beta_k \left( 
		\log n_k - 1 
	\right)
\end{align*}

Taking the inner product with $q$ gives
\begin{align*}
	\langle \nabla d(q), q \rangle
	= \sum_{j\in S_Q; i\in \bbI_j} q_i \left[ \beta_j(\log \frac{q_{i}}{q_{p_j}} + 1) 
	+ \sum_{k \in \cD_j^i} \beta_k \left( 
		\log n_k - 1 
	\right)\right]
\end{align*}

Subtracting $-d(q)$ gives
\begin{align*}
	-d(q) + \langle \nabla d(q), q \rangle 
	= -&   \sum_{j\in S_Q; i\in \bbI_j} \beta_j q_i \log\frac{q_i}{q_{p_j}} 
	- \sum_{j\in S_Q} \beta_j q_{p_j} \log(n_j)\\
	+& \sum_{j\in S_Q; i\in \bbI_j} q_i \left[ \beta_j(\log \frac{q_{i}}{q_{p_j}} + 1) 
	+ \sum_{k \in \cD_j^i} \beta_k \left( 
		\log n_k - 1 
	\right)\right] \\
	= & \sum_{j\in S_Q; i\in \bbI_j} q_i \beta_j
	+ \sum_{j\in S_Q}\sum_{k \in \cD_j^i} q_{p_k} \beta_k \left( 
		\log n_k - 1 
	\right)
	- \sum_{j\in S_Q} \beta_j q_{p_j} \log(n_j) \\
	= & \sum_{j\in S_Q} q_{p_j} \beta_j
	+ \sum_{j\in S_Q}\sum_{k \in \cD_j^i} q_{p_k} \beta_k \left( 
		\log n_k - 1 
	\right)
	- \sum_{j\in S_Q} \beta_j q_{p_j} \log(n_j) \\
	= & 
	\sum_{j\in S_Q}\sum_{k \in \cD_j^i} q_{p_k} \beta_k \left( 
		\log n_k - 1 
	\right)
	- \sum_{j\in S_Q} \beta_j q_{p_j} (\log(n_j) - 1) \\
	= & 
	- \sum_{j\in S_Q; b_Q^j = 0} \beta_j q_{p_j} (\log(n_j) - 1)
\end{align*}

%%% Local Variables:
%%% mode: latex
%%% TeX-master: "../nips1818/gpu_egt_river_nips1818"
%%% End:

\end{document}